\documentclass{emulateapj}

\shorttitle{First Neutrino Point-Source Results from the 22-String IceCube Detector}
\shortauthors{IceCube Collaboration: R.~Abbasi et al.}

\usepackage{graphicx}
\usepackage{dcolumn}
\usepackage{bm}

\begin{document}

\title{First Neutrino Point-Source Results from the 22-String IceCube Detector}

\author{
IceCube Collaboration:
R.~Abbasi\altaffilmark{1},
Y.~Abdou\altaffilmark{2},
M.~Ackermann\altaffilmark{3},
J.~Adams\altaffilmark{4},
J.~Aguilar\altaffilmark{1},
M.~Ahlers\altaffilmark{5},
K.~Andeen\altaffilmark{1},
J.~Auffenberg\altaffilmark{6},
X.~Bai\altaffilmark{7},
M.~Baker\altaffilmark{1}
% ...
}
\altaffiltext{1}{Dept.~of Physics, University of Wisconsin, Madison, WI 53706, 
USA; cfinley@icecube.wisc.edu, jdumm@icecube.wisc.edu
}
\altaffiltext{2}{Dept.~of Subatomic and Radiation Physics, University of Gent, B-9000 Gent, Belgium}
\altaffiltext{3}{DESY, D-15735 Zeuthen, Germany}
\altaffiltext{4}{Dept.~of Physics and Astronomy, University of Canterbury, Private Bag 4800, Christchurch, New Zealand}
\altaffiltext{5}{Dept.~of Physics, University of Oxford, 1 Keble Road, Oxford OX1 3NP, UK}
\altaffiltext{6}{Dept.~of Physics, University of Wuppertal, D-42119 Wuppertal, Germany}
\altaffiltext{7}{Bartol Research Institute and Department of Physics and Astronomy, University of Delaware, Newark, DE 19716, USA}
% ...


\begin{abstract}

We present new results of searches for neutrino point sources in the northern sky, using data recorded in 2007-08 with 22 strings of the IceCube detector (approximately one-fourth of the planned total) and 275.7 days of livetime. The final sample of 5114 neutrino candidate events agrees well with the expected background of atmospheric muon neutrinos and a small component of atmospheric muons.  No evidence of a point source is found, with the most significant excess of events in the sky 
at $2.2\,\sigma$ after accounting for all trials.  The average upper limit over the northern sky for point sources of muon-neutrinos with $E^{-2}$ spectrum is $E^{2}\,\Phi_{\nu_{\mu}} < 1.4 \times 10^{-11}\, \mathrm{TeV\,cm}^{-2} \mathrm{\,s}^{-1}$, in the energy range from 3\,TeV to 3\,PeV, improving the previous best average upper limit by the AMANDA-II detector by a factor of two.

\end{abstract}


\section{\label{sec:introduction}Introduction}

Cosmic rays with energies up to $10^{20}$\,eV pervade the Universe, but their sources remain unknown.  Possible acceleration sites of cosmic rays include shock fronts in supernova remnants, pulsars, microquasars, active galactic nuclei, and gamma-ray bursts.
While many of these sources are now observed by gamma-ray astronomy experiments \citep{Milagro_Abdo:2007ad, HESS_Aharonian:2005kn, MAGIC_microquasars_Albert:2006vk}, 
it remains difficult to determine whether the gamma-ray emission is of leptonic or hadronic origin.  
Hadronic acceleration in these sources is expected to produce a correlated neutrino flux as accelerated protons interact with ambient gas and radiation to produce mesons, and the charged mesons decay to neutrinos
 (for reviews see 
\citet{Becker:2007sv,
Bednarek:2004ky,
Halzen:2002pg} and references therein).
This signature uniquely distinguishes hadronic from leptonic processes, and thus detection of the high-energy neutrino flux from cosmic ray accelerators is the key to identifying them.  Moreover, since neutrinos can propagate freely through dense environments and across cosmological distances that are optically thick to photons, they can probe hidden regions and reveal unexpected sources, opening a unique window on the high-energy Universe.

Previous searches for high-energy astrophysical neutrino sources performed by MACRO \citep{MACRO_Ambrosio:2000yx}, Super-Kamiokande \citep{SuperK_Desai:2007ra}, and AMANDA \citep{AMANDA7YR_Collaboration:2008ih} have set upper limits that demonstrate the need for much larger experiments to detect these weakly interacting particles.  The IceCube Neutrino Observatory is now instrumenting a cubic kilometer of the clear Antarctic ice sheet at the geographical South Pole.
Construction began in the austral summer 2004--05, and is planned to finish in 2011.
The full detector will comprise 4,800 Digital Optical Modules (DOMs) deployed on 80 strings between 1.5--2.5\,km deep within the ice, a surface array (IceTop) for observing extensive air showers of cosmic rays, and an additional dense subarray (DeepCore) in the detector center for enhanced low-energy sensitivity.  Each DOM consists of a 25\,cm diameter Hamamatsu photo-multiplier tube, electronics for waveform digitization \citep{IceCubeDAQ_:2008ym}, and a spherical, pressure-resistant glass housing.
Waveforms are recorded when nearest or next-to-nearest DOMs fire within $\pm\,1$\,microsecond; event triggers occur when eight DOMs record waveforms within 5\,microseconds.
  Calibrations ensuring nanosecond timing precision are described in \citet{IceCubeFirstYearPerformance_Achterberg:2006md}.  The DOMs detect Cherenkov photons emitted by relativistic charged particles passing through the ice.  In particular, the directions of muons (either from cosmic ray showers above the surface, or neutrino interactions within the ice or bedrock) can be well reconstructed from the track-like pattern and timing of hit DOMs.
Identification of neutrino-induced muon events in IceCube has been demonstrated in \citet{IceCube9StringAtmNu_Achterberg:2007bi} using atmospheric neutrinos as a calibration tool.

\section{Event Selection and Analysis}

As of spring 2007, there were 22 deployed IceCube strings.  The physics run for the 22-string configuration started 2007 May 31 and ended 2008 April 4, when the 40-string configuration began operating.  The final livetime is 275.7 days, about 90\% of the total available time including operation during the construction season.  The event trigger rate is $\sim 550$\,Hz, predominantly due to down-going muons.  The rate of atmospheric neutrino-induced muons triggering the detector is roughly $10^{6}$ times lower.  Because only neutrino-induced muons can travel upwards, neutrino events can be isolated by selecting up-going tracks.  An online event filter makes the first rejection of down-going tracks.  Events which pass (at a rate $\sim 20$\,Hz) are sent over satellite to the North.  Likelihood-based track reconstructions are performed, improving the directional accuracy and background rejection capabilities as well as providing individual angular uncertainty estimates \citep{Paraboloid_Neunhoffer:2004ha}.  The reduced log-likelihood of the best-fit track, the angular uncertainty, and the number of modules which were hit by direct Cherenkov photons (within a window of $-15$ to $+75$\,ns for estimated hit times of the reconstructed track) are the main parameters used to select up-going neutrino candidates and reject background.  An additional cut on the likelihood ratio of the best-fit track to the best-fit track constrained to be down-going further reduces background close to the horizon.  In the final analysis, a wide range of cuts based on these parameters are compatible with optimal sensitivity to both hard and soft spectrum sources.  Within this range, applying cuts that also remove the largest fraction of mis-reconstructed down-going events yields a final sample consisting of 5114 neutrino candidate events.

At this time there is not a single strategy for point source searches in IceCube, so more than one approach was investigated, including a binned analysis similar to that in \citet{AMANDA5YR_Achterberg:2006vc}, as well as an unbinned analysis similar to that in \citet{AMANDA7YR_Collaboration:2008ih}.  Simulation studies using the 22-string IceCube configuration showed that the latter approach was on average 35\% more sensitive for both hard and soft point source spectra.  For this reason, it was decided before the data were unblinded that the unbinned analysis would be used for the final results, which are reported below.

The unbinned likelihood analysis is described in detail in \citet{Braun:2008bg_Methods}; it uses both the direction and energy information of each event.  Astrophysical neutrino source spectra are typically expected to be harder ($\sim\,E^{-2}$ in the Fermi model of cosmic ray acceleration) than the known spectrum of the atmospheric neutrino background ($\sim\,E^{-3.7}$).  Thus a neutrino point source may be detectable not just by the clustering of event arrival directions, but by a different event energy distribution than the background.  For each direction in the sky tested, the analysis performs a fit for the number of signal events $n_{s}$ above background, and the spectral index $\gamma$ of the excess events.  The test-statistic in the analysis is the log likelihood ratio of the signal hypothesis with best-fit parameters ($\hat{n}_{s}$ and $\hat{\gamma}$) to the null hypothesis of no signal present ($n_{s}=0$).   This test-statistic provides an estimate of the significance (pre-trial p-value) of deviation from background at a given position in the sky.  As described below, the post-trial significance is determined by applying the analysis to scrambled data sets, in which the right ascension of the events are randomized but all other event properties are kept the same.

Two unbinned point-source searches are performed.  The first is an all-sky search within the declination range $-5^{\circ}$ to $+85^{\circ}$: the maximum likelihood ratio is evaluated for each direction in the sky in steps of $0.25^{\circ}$ r.a.\ and $0.25^{\circ}$ dec., (well below the angular resolution of $1.5^{\circ}$).  The significance of any spot is given by the fraction of scrambled data sets containing at least one spot with a log likelihood ratio higher than the one observed in the real data.  This fraction is the post-trial p-value.  Because the all-sky search involves a large number of effective trials, the second search is restricted to the directions of 28 {\it a priori} selected source candidates, in order to improve the confidence of a possible detection of one of these objects.  The post-trial p-value is again found by performing the source list analysis on scrambled data sets.  The smallest post-trial p-value from either of the two searches is then taken as the final significance of the analysis, with a final trial factor of two.

\section{Detector Response}

\begin{figure}
\includegraphics[width=0.5\textwidth]{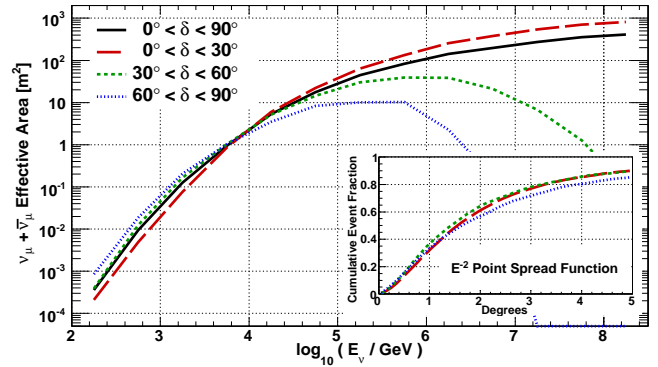}
\caption{\label{fig:eff}
Solid-angle-averaged effective areas at final cut level for astrophysical muon neutrino fluxes ($\nu_{\mu} + \bar{\nu}_{\mu}$) at different declinations. The turnover at high-energy for up-going events is due to absorption by Earth.  Inset: $E^{-2}$ point spread function (angular difference between the neutrino and reconstructed muon track) for the same declination ranges.}
\end{figure}

\begin{figure}
\includegraphics[width=0.5\textwidth]{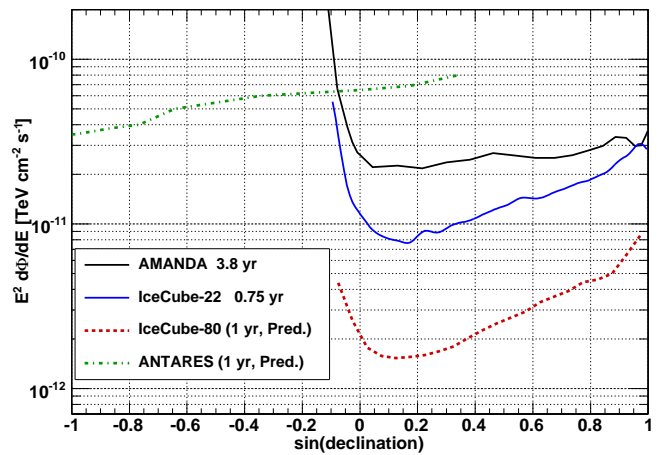}
\caption{\label{fig:sens}
Sensitivity to a point-source $E^{-2}\,\nu_{\mu}$ flux as a function of declination, shown for the final AMANDA-II analysis \citep{AMANDA7YR_Collaboration:2008ih}, the current twenty-two string IceCube analysis, and the predicted sensitivities for the ANTARES experiment \citep{AguilarSanchez:2007hh} and the final IceCube configuration.}
\end{figure}

A simulation of $\nu_{\mu}$ and $\bar{\nu}_{\mu}$ was used to determine the effective area and point spread function for the 22-string IceCube configuration, shown in Fig. ~\ref{fig:eff}.  The sky-averaged median angular reconstruction error is $1.5^{\circ}$ for both $E^{-2}$ and atmospheric spectra.  For an $E^{-2}$ (atmospheric) neutrino spectrum, 90\% of the events are in the central range 3\,TeV -- 3\,PeV (250\,GeV -- 16\,TeV).  The sensitivity (median upper limit, following the ordering principle of \citet{Feldman:1997qc}) as a function of declination is shown in Fig.~\ref{fig:sens}.  For a source at declination $5^{\circ}$ with $E^{-2}$ spectrum, approximately 13 (16) signal events are needed for a 50\% chance of a post-trial 5\,$\sigma$ detection based on the source list search (all sky search).

Systematic uncertainties arise chiefly from the modeling of light propagation in the ice.  Simulation of muons passing through the ice is described in \citet{Chirkin:2004hz}, and the propagation of light from the muon to the optical modules is performed with Photonics \citep{Lundberg:2007mf}.  
Embedded impurities in the ice alter the scattering and absorption properties with a depth dependence due to climate changes over tens of thousands of years; direct measurements of these ice properties are used in the detector simulation \citep{OpticalProperties_JGeophysRes_2006}.  Data--Monte Carlo comparisons using both trigger level and high-quality down-going muons indicate the systematic uncertainty which remains from this aspect of simulation is 15\%.  Systematic uncertainties on the optical module efficiency and true point spread function contribute an additional 9\% uncertainty on neutrino flux estimates, while theoretical uncertainties on muon energy losses and the neutrino-nucleon cross-section \citep{CTEQ6_Pumplin:2002vw} contribute 5\%.  The total estimated systematic uncertainty of 18\% has been incorporated into the sensitivity and upper limit calculations using the methods described in \citet{Conrad:2002kn} and \citet{Hill:2003jk}.  Because taus from tau neutrinos also decay to muons (with a 17\% branching ratio), any assumption of an additional component of tau neutrinos in the astrophysical flux arriving at earth would lead to a tighter upper limit on the muon neutrino flux.

\begin{figure}
\includegraphics[width=0.5\textwidth]{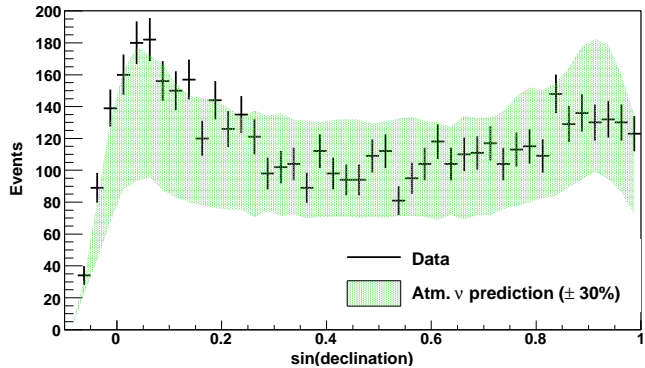}
\caption{\label{fig:atmNu}
Declination distribution of the final data sample and simulated atmospheric neutrino events with 30\% theoretical uncertainty on the absolute normalization of the predicted neutrino flux \citep{Barr:2004br, BartolUncertainty_Barr:2006it}.}
\end{figure}

Atmospheric neutrino events, while posing a substantial background to extra-terrestrial neutrino searches, provide a useful verification of the detector and simulation.  In Fig.~\ref{fig:atmNu}, the declination distribution is shown for the data and atmospheric neutrino simulation with final cuts applied. 
The atmospheric neutrino model has a theoretical uncertainty of 30\% in the flux normalization in the TeV energy range \citep{Barr:2004br, BartolUncertainty_Barr:2006it, IceCube9StringAtmNu_Achterberg:2007bi}.  This uncertainty does not affect astrophysical flux calculations, but limits the precision of atmospheric neutrinos as a check of the detector simulation.  In 275.7 days of livetime, $4600 \pm 1400$ atmospheric neutrinos are expected, along with an additional component of $400 \pm 200$ mis-reconstructed down-going muons from air showers (simulated with CORSIKA \citep{Heck:1998vt}),
mainly near the horizon.  Within uncertainties, this is in agreement with the 5114 events observed.

\section{Results}

\begin{figure*}
\includegraphics[width=1.0\textwidth]{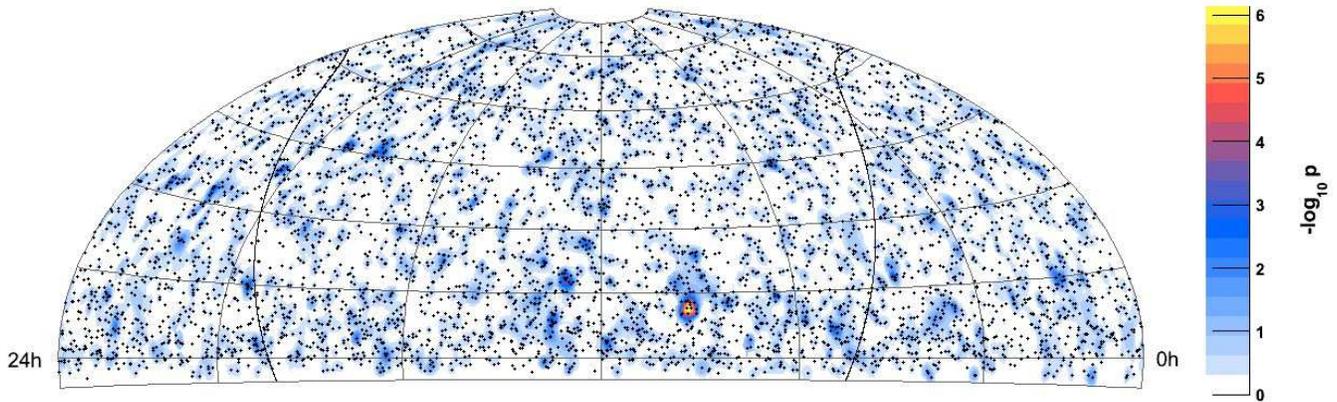}
\caption{\label{fig:AllSky}
Equatorial skymap of events (points) and pre-trial significances (p-value) of the all-sky point source search.  The solid curve is the galactic plane.}
\end{figure*}

The results of the all-sky search are shown in Fig.~\ref{fig:AllSky}.  The most significant deviation from background is located at $153.4^{\circ}$ r.a., $11.4^{\circ}$ dec.  The best-fit parameters are $n_{s}= 7.7$ signal events above background, with spectral index $\gamma = -1.65$.  The pre-trial estimated p-value of the maximum log likelihood ratio at this location is $7\times10^{-7}$.  The post-trial p-value is determined by performing the analysis with the right ascension of the data randomized: 67 out of 10,000 scrambled data sets yielded a more significant excess somewhere in the sky.

The results of the point-source search in the direction of 28 source candidates selected {\it a priori} are given in Table~\ref{tab:sourcelist}.  The smallest pre-trial estimated p-value is 7\% (for the TeV blazar 1ES\,1959+650); 66 out of 100 scrambled data sets have a more significant excess for at least one source on this list.

\begin{table}
\caption{\label{tab:sourcelist}
Results for the {\it a priori} source candidate list}
\begin{ruledtabular}  
\begin{tabular}{ccccccc}
Object & r.a.~$[^{\circ}]$ & dec.~$[^{\circ}]$ &  $n_{s}$ & $p$ & $\Phi_{90}$ & $B_{2^{\circ}}$ \\
\hline
       MGRO J2019+37 & 304.83 & 36.83 & 3.1 & 0.25 & 25.23 & 2.6\\
       MGRO J1908+06 & 287.27 &  6.28 & 0.0 & -- &  7.06 & 3.7\\
             Cyg OB2 & 308.08 & 41.51 & 0.0 & -- & 15.28 & 2.6\\
              SS 433 & 287.96 &  4.98 & 2.8 & 0.32 & 11.65 & 4.1\\
             Cyg X-1 & 299.59 & 35.20 & 0.0 & -- & 14.60 & 2.4\\
        LS I +61 303 &  40.13 & 61.23 & 0.0 & -- & 22.00 & 3.0\\
        GRS 1915+105 & 288.80 & 10.95 & 0.0 & -- &  7.64 & 3.3\\
       XTE J1118+480 & 169.54 & 48.04 & 2.5 & 0.082 & 40.62 & 2.8\\
        GRO J0422+32 &  65.43 & 32.91 & 0.0 & -- & 14.10 & 2.2\\
             Geminga &  98.48 & 17.77 & 0.0 & -- &  9.67 & 2.6\\
         Crab Nebula &  83.63 & 22.01 & 0.0 & -- & 10.35 & 2.4\\
               Cas A & 350.85 & 58.81 & 0.0 & -- & 20.22 & 3.5\\
             Mrk 421 & 166.11 & 38.21 & 0.0 & -- & 14.35 & 2.8\\
             Mrk 501 & 253.47 & 39.76 & 0.0 & -- & 14.44 & 2.7\\
        1ES 1959+650 & 300.00 & 65.15 & 5.0 & 0.071 & 59.00 & 3.2\\
        1ES 2344+514 & 356.77 & 51.70 & 0.0 & -- & 17.94 & 2.8\\
          H 1426+428 & 217.14 & 42.67 & 0.0 & -- & 15.64 & 2.7\\
        1ES 0229+200 &  38.20 & 20.29 & 0.0 & -- & 10.24 & 2.4\\
              BL Lac & 330.68 & 42.28 & 1.6 & 0.37 & 22.81 & 2.7\\
          S5 0716+71 & 110.47 & 71.34 & 1.9 & 0.31 & 44.76 & 3.3\\
               3C66A &  35.67 & 43.03 & 2.0 & 0.31 & 25.70 & 2.8\\
            3C 454.3 & 343.49 & 16.15 & 0.0 & -- &  9.07 & 2.6\\
            4C 38.41 & 248.82 & 38.13 & 0.0 & -- & 14.29 & 2.8\\
        PKS 0528+134 &  82.74 & 13.53 & 0.0 & -- &  8.27 & 3.2\\
              3C 273 & 187.28 &  2.05 & 0.9 & 0.37 & 11.73 & 4.1\\
                 M87 & 187.71 & 12.39 & 0.0 & -- &  7.91 & 3.2\\
            NGC 1275 &  49.95 & 41.51 & 2.2 & 0.21 & 28.32 & 2.6\\
               Cyg A & 299.87 & 40.73 & 0.0 & -- & 15.05 & 2.6\\
\end{tabular}
\end{ruledtabular}   
\tablecomments{$n_{s}$ is the best-fit number of signal events; when $n_{s}>0$ the (pre-trial) p-value is also calculated.  $\Phi_{90}$ is the upper limit of the Feldman-Cousins 90\% confidence interval for an $E^{-2}$ flux, i.e.: $d\Phi/dE \leq \Phi_{90} \, 10^{-12}\mathrm{TeV}^{-1} \mathrm{cm}^{-2} \mathrm{s}^{-1} (E / \mathrm{TeV})^{-2}$.  The background event density at the source declination is indicated by the mean number of background events $B_{2^{\circ}}$ expected in a bin of radius $2^{\circ}$.  }
\end{table}

Of the two searches, the most significant result comes from the all-sky search.  Accounting for this last trial factor of two, the final p-value for the analysis is 1.34\%.  At this level of significance, the excess is consistent with the background-only null hypothesis.  If not a statistical fluctuation, the excess will be detectable with future IceCube data, unless it were caused by a one-time or rare astronomical event.  Subsequent examination of the times of the events in the region of excess, however, has not revealed any burst-like distribution in time, with the ten events that contribute most to the excess distributed throughout the year and each separated by a minimum of nine days from the next.

\section{Conclusions}

A search for point sources of high-energy neutrinos has been performed using data recorded during 2007-08 with 22 strings of IceCube.  An all-sky search within the declination range $-5^{\circ}$ to $+85^{\circ}$ found the most significant deviation from the background at $153.4^{\circ}$ r.a., $11.4^{\circ}$ dec.  Accounting for all trials in the point source search, the final p-value for this result is 1.34\%, consistent with the null hypothesis of background-only events at the $2.2~\sigma$ level (if the p-value is expressed as the one-sided tail of a Gaussian distribution).  No obvious source candidates are near this location, and an analysis of the timing of the events did not find any evidence of a burst in time.  The location can be added to the {\it a priori} source candidate list for analysis using future IceCube data, in which case a similar excess would be identified with much higher significance.

No evidence of neutrino emission was found for any of the 28 {\it a priori} selected point-source candidates, and the resulting upper limits severely constrain some models of neutrino emission.  For example, simulation of the neutrino flux from the microquasar SS433, using the model of \citet{Distefano:2002qw} and expressed as a broken power-law \citep{Bednarek:2004ky} with a cutoff at 100 TeV, predicts a point-source signal of 48 neutrino events in the 22-string data sample, and is now well excluded.  Alternative predictions for SS433 \citep[e.g.][]{Reynoso:2008nk} will however only be testable with the full IceCube detector.  Correlation of the neutrino arrival times with the known orbital period of the binary system or the precession period of the jets may enhance the sensitivity to such objects and provide insight into the emission processes. 

The sensitivity of this search with one season of 22-string data already exceeds the combined sensitivity of all previous neutrino point-source searches in the TeV-PeV energy range. New searches are underway to extend the sensitivity to ultrahigh-energy sources in the southern sky, and to lower energy sources using events recorded by the combined IceCube-AMANDA detector.  With completion of the full 80-string detector expected in 2011, the improved acceptance, signal efficiency, background rejection and angular resolution ($< 0.8^{\circ}$) should provide more than an order of magnitude enhancement in sensitivity within several years of operation.

\acknowledgments

We acknowledge the support from the following agencies:
U.S. National Science Foundation-Office of Polar Program,
U.S. National Science Foundation-Physics Division,
University of Wisconsin Alumni Research Foundation,
U.S. Department of Energy, and National Energy Research Scientific Computing Center,
the Louisiana Optical Network Initiative (LONI) grid computing resources;
Swedish Research Council,
Swedish Polar Research Secretariat,
and Knut and Alice Wallenberg Foundation, Sweden;
German Ministry for Education and Research (BMBF),
Deutsche Forschungsgemeinschaft (DFG), Germany;
Fund for Scientific Research (FNRS-FWO),
Flanders Institute to encourage scientific and technological research in industry (IWT),
Belgian Federal Science Policy Office (Belspo);
the Netherlands Organisation for Scientific Research (NWO);
M.~Ribordy acknowledges the support of the SNF (Switzerland);
A.~Kappes and A.~Gro{\ss} acknowledge support by the EU Marie Curie OIF Program;
J.~P.~Rodrigues acknowledge support by the Capes Foundation, Ministry of Education of Brazil.

\bibliographystyle{apj}
\bibliography{ic22ps}

\end{document}